\documentclass[AMA,STIX1COL]{WileyNJD-v2}
\usepackage{moreverb}
\usepackage{csquotes}
\usepackage{graphicx,color}
\usepackage{float}
\usepackage[caption=false]{subfig}
\usepackage{amsmath}
\usepackage{amsmath}
\usepackage{multirow}
\usepackage{dsfont}
\usepackage[shortlabels]{enumitem}
\usepackage{epstopdf}
\newcommand\BibTeX{{\rmfamily B\kern-.05em \textsc{i\kern-.025em b}\kern-.08em
T\kern-.1667em\lower.7ex\hbox{E}\kern-.125emX}}
\articletype{ORIGINAL ARTICLE}%
\received{<day> <Month>, <year>}
\revised{<day> <Month>, <year>}
\accepted{<day> <Month>, <year>}
\begin{document}
\title{On the density--density correlations of the non-interacting finite temperature electron gas}
\author{Panagiotis Tolias$^1$}
\author{Tobias Dornheim$^{2,3}$}
\author{Jan Vorberger$^3$}
\address{$^1$\orgdiv{Space and Plasma Physics}, \orgname{Royal Institute of Technology (KTH)}, \orgaddress{\state{Stockholm}, \country{Sweden}}}
\address{$^2$\orgname{Center for Advanced Systems Understanding (CASUS)}, \orgaddress{\state{G\"orlitz}, \country{Germany}}}
\address{$^3$\orgname{Helmholtz-Zentrum Dresden-Rossendorf (HZDR)}, \orgaddress{\state{Dresden}, \country{Germany}}}
\authormark{P. Tolias}
\corres{Panagiotis Tolias, Space and Plasma Physics, Royal Institute of Technology, Stockholm SE-100 44, Sweden. \email{tolias@kth.se}}
\abstract[Abstract]{The density--density correlations of the non-interacting finite temperature electron gas are discussed in detail. Starting from the ideal linear density response function and utilizing general relations from linear response theory, known and novel expressions are derived for the pair correlation function, static structure factor, dynamic structure factor, thermal structure factor and imaginary time correlation function. Applications of these expressions in the classical mapping approach, self-consistent dielectric formalism and equation-of-state construction are analyzed in depth.}
\jnlcitation{\cname{%
\author{P. Tolias, T. Dornheim and J. Vorberger}} (\cyear{2024}),
\ctitle{On the density--density correlations of the non-interacting finite temperature electron gas}, \cjournal{Contributions to Plasma Physics}.}
\maketitle

\section{Introduction}\label{sec:intro}

The non-interacting limit of any classical or quantum many body system holds a special role in its mathematical description in most theoretical frameworks; see the plasma fluctuation theory of Klimontovich\cite{Klimontovich1958,Klimontovich1967}, the (non-)equilibrium Green's function formalism\cite{KrempBook,BonitzGBook}, the Fermi liquid theory of Landau\cite{Abrikosov1959,PinesIBook} or the local effective field concept of linear response theory\cite{GiulianiVignale,IchimaruBook}. From the early days of many body theory, the free or ideal linear density response of the electron gas was computed either from the equation of motion of $N$ independent electrons\cite{GiulianiVignale} or, directly in the thermodynamic limit, from one of the many representations of quantum kinetic theory\cite{KlimontovichUsp1960}. The deep understanding of the properties of the linear density response function of the non-interacting electron gas is rather mandatory for practitioners in the fields of condensed matter physics and high energy density physics. In fact, standard introductory textbooks in these fields typically dedicate separate sections to the so-called Lindhard density response for arbitrary dimensionality\cite{GiulianiVignale,MahanBook}. Nevertheless, the discussion is usually confined to the analyticity properties and the branch cut along the real axis as well as the long wavelength limit and the static limit, often barely touching upon the fact that the density-density correlations of the non-interacting electron gas are non-trivial as a consequence of exchange effects. In addition, the discussion is often restricted to the ground state and the high temperature classical limit, where closed form expressions are available for most properties, and an in depth analysis of the general case of arbitrary degeneracy is lacking.

In this pedagogical article, we analyze the whole range of density-density correlation functions of the non-interacting finite temperature electron gas. The focus lies on the thermodynamically stable paramagnetic case and on the warm dense matter regime, where neither the ground state limit nor the classical limit are adequate. Starting from the Lindhard density response and utilizing general relations from linear response theory, we derive known and novel expressions for the pair correlation function, the static structure factor, the dynamic structure factor, the thermal structure factor and the imaginary time correlation function of the non-interacting electron gas. We discuss elementary applications of these expressions in the classical mapping approach\cite{DharmawardanaPRL2000}, the self-consistent dielectric formalism\cite{IchimaruPhysRep1987} and the parameterization of the uniform electron gas equation of state\cite{DornheimPhysRep2018}. This article sets the stage for future investigations of the high order density correlations of the non-interacting finite temperature electron gas that are invariably connected with the nonlinear density response\cite{vorberger2024greensfunctionperspectivenonlinear}.

\section{Theoretical background}\label{sec:background}

\subsection{Density--density correlation functions}\label{sec:backgroundCor}

It is instructive to introduce the entire range of spatiotemporal density--density correlation functions. The microscopic number density operator is naturally given by $\hat{\rho}(\boldsymbol{r})=\sum_{i=1}^{N}\delta(\boldsymbol{r}-\hat{\boldsymbol{r}}_i)$ in real space and by $\hat{\rho}_{\boldsymbol{k}}=\sum_{i=1}^{N}e^{-\imath\boldsymbol{k}\cdot\hat{\boldsymbol{r}}_i}$ in reciprocal space, with $N$ the particle number. The starting point is the \emph{van Hove function} that is defined in real space as\cite{HansenBook}
\begin{align}
G(\boldsymbol{r},t)=\frac{1}{N}\int\langle\hat{\rho}(\boldsymbol{r}^{\prime}+\boldsymbol{r},t)\hat{\rho}(\boldsymbol{r}^{\prime},0)\rangle_0d^3r^{\prime}\,.\label{eq:vanHovedef}
\end{align}
The \emph{intermediate scattering function} (ISF) is defined as the spatial Fourier transform of the van Hove function, i.e.,\cite{HansenBook}
\begin{align}
F(\boldsymbol{k},t)=\int\,G(\boldsymbol{r},t)e^{-\imath\boldsymbol{k}\cdot\boldsymbol{r}}d^3r=\frac{1}{N}\left\langle\hat{\rho}_{\boldsymbol{k}}(t)\hat{\rho}^{\dagger}_{\boldsymbol{k}}\right\rangle_0\,.\label{eq:ISFdef}
\end{align}
The \emph{dynamic structure factor} (DSF) is defined as the power spectrum of the ISF, i.e.,\cite{HansenBook,SheffieldBook}
\begin{align}
S(\boldsymbol{k},\omega)=\frac{1}{2\pi}\int_{-\infty}^{+\infty}F(\boldsymbol{k},t)e^{\imath\omega{t}}dt=\frac{1}{N}\frac{1}{2\pi}\int_{-\infty}^{+\infty}\left\langle\hat{\rho}_{\boldsymbol{k}}(t)\hat{\rho}^{\dagger}_{\boldsymbol{k}}\right\rangle_0e^{\imath\omega{t}}dt\,.\label{eq:DSFdef}
\end{align}
The \emph{imaginary time density-density correlation function} or imaginary time correlation function (ITCF) is defined by setting $t=-\imath\hbar\tau$ to the ISF. This correlation function, when $\tau\in[0,\beta]$ where $\beta=1/T$ with $T$ the temperature in energy units, can be directly extracted from path-integral Monte Carlo (PIMC) simulations of finite temperature systems\cite{CeperleyRMP1995}. It is also straightforward to prove that the ITCF is the two-sided Laplace transform of the DSF, i.e.,\cite{CeperleyRMP1995,JarrellPhysRep1996,DornheimPRB2023}
\begin{align}
F(\boldsymbol{k},\tau)=\int_{-\infty}^{+\infty}S(\boldsymbol{k},\omega)e^{-\hbar\omega\tau}d\omega=\frac{1}{N}\left\langle\hat{\rho}_{\boldsymbol{k}}(-\imath\hbar\tau)\hat{\rho}^{\dagger}_{\boldsymbol{k}}\right\rangle_0\,.\label{eq:ITCFdef}
\end{align}
The \emph{static structure factor} (SSF) is the normalization of the DSF as well as the initial value of the ISF and the ITCF, i.e.,\cite{GiulianiVignale,HansenBook}
\begin{align}
S(\boldsymbol{k})=\int_{-\infty}^{+\infty}S(\boldsymbol{k},\omega)d\omega=\frac{1}{N}\left\langle\hat{\rho}_{\boldsymbol{k}}\hat{\rho}^{\dagger}_{\boldsymbol{k}}\right\rangle_0=F(\boldsymbol{k},t=0)=F(\boldsymbol{k},\tau=0)\,.\label{eq:SSFdef}
\end{align}
The \emph{thermal structure factor} (TSF) is defined as another special value of the ITCF, now at $\tau=\beta/2$, i.e.,\cite{DornheimPoPRev2023,DornheimJCP2023}
\begin{align}
S^{\beta/2}(\boldsymbol{k})=\frac{1}{N}\left\langle\hat{\rho}_{\boldsymbol{k}}(-\imath\hbar\beta/2)\hat{\rho}^{\dagger}_{\boldsymbol{k}}\right\rangle_0=F(\boldsymbol{k},\tau=\beta/2)\,.\label{eq:TSFdef}
\end{align}
Finally, the \emph{pair correlation function} (PCF) can be introduced by following the hierarchy of the reduced s-particle density matrices, the reduced s-particle densities and ultimately the reduced s-particle correlation functions\cite{BonitzBook}. For homogeneous systems, it is defined by\cite{HansenBook,McQuarrieBook}
\begin{align}
ng(\boldsymbol{r})=\frac{1}{N}\left\langle\sum_{i=1}^{N}\sum_{j\neq{i}}^{N}\delta(\boldsymbol{r}+\hat{\boldsymbol{r}}_j-\hat{\boldsymbol{r}}_i)\right\rangle_0\,.\label{eq:PFCdef}
\end{align}
In contrast to the PCF, the \emph{total correlation function} (TCF) $h(\boldsymbol{r})=g(\boldsymbol{r})-1$ is an absolutely integrable function that possesses a well-behaved spatial Fourier transform. Being an equal-time density correlation function, it is unsurprising that the TCF is connected to the SSF via the simple expression\cite{HansenBook,McQuarrieBook}
\begin{align}
S(\boldsymbol{k})=1+nH(\boldsymbol{k})\,,\label{eq:microconnection}
\end{align}
where $n$ is the particle density and $H(\boldsymbol{k})$ is the Fourier transform of the TCF.

\subsection{Linear density response theory}\label{sec:backgroundLRT}

Let us recap some basic definitions and relations from the density-density version of linear response theory\cite{GiulianiVignale,IchimaruBook,KuboBook}. The density--density (or simply density) response function $\chi_{\mathrm{nn}}(\boldsymbol{r},\boldsymbol{r}^{\prime},t)\equiv\chi_{n(\boldsymbol{r})n(\boldsymbol{r}^{\prime})}(t)$ describes the $\delta{n}(\boldsymbol{r},t)$ change in the expectation value of the microscopic number density operator $\hat{\rho}(\boldsymbol{r})$ at point $\boldsymbol{r}$ and time instant $t$ under the influence of an applied potential energy $V_{\mathrm{ext}}(\boldsymbol{r}^{\prime},t^{\prime})$ that couples linearly to the microscopic number density operator $\hat{\rho}(\boldsymbol{r}^{\prime})$ at point $\boldsymbol{r}^{\prime}$. Thus, the additional term to the unperturbed Hamiltonian operator is simply given by $\int{V}_{\mathrm{ext}}(\boldsymbol{r}^{\prime},t^{\prime})\hat{\rho}(\boldsymbol{r}^{\prime})d^3r^{\prime}$. In real space, the density response function is formally defined by the functional derivative\cite{GiulianiVignale,IchimaruBook}
\begin{align}
\chi_{\mathrm{nn}}(\boldsymbol{r},\boldsymbol{r}^{\prime},t-t^{\prime})=\left.\frac{\delta{n}(\boldsymbol{r},t)}{\delta{V}_{\mathrm{ext}}(\boldsymbol{r}^{\prime},t^{\prime})}\right|_{V_{\mathrm{ext}}=0}\,.\label{eq:DRFdef}
\end{align}
with the respective real space Kubo formula reading as\cite{GiulianiVignale}
\begin{align}
\chi_{\mathrm{nn}}(\boldsymbol{r},\boldsymbol{r}^{\prime},t)=-\frac{\imath}{\hbar}\mathrm{H}(t)\langle[\hat{\rho}(\boldsymbol{r},t),\hat{\rho}(\boldsymbol{r}^{\prime})]\rangle_0\,,\label{eq:DRFkubo}
\end{align}
where $\mathrm{H}(\cdot)$ denotes the Heaviside step function. In reciprocal space, one has $\chi_{\mathrm{nn}}(\boldsymbol{k},\boldsymbol{k}^{\prime},\omega)=(1/V)\chi_{\hat{\rho}_{\boldsymbol{k}}\hat{\rho}_{-\boldsymbol{k}^{\prime}}}(\omega)$ with $V$ the volume. For homogeneous systems, one also has $\chi_{\mathrm{nn}}(\boldsymbol{k},\boldsymbol{k}^{\prime},\omega)=(1/V)\delta(\boldsymbol{k}-\boldsymbol{k}^{\prime})\chi_{\hat{\rho}_{\boldsymbol{k}}\hat{\rho}_{\boldsymbol{k}^{\dagger}}}(\omega)$. This allows to introduce the two-argument version $\chi(\boldsymbol{k},\omega)=(1/V)\chi_{\hat{\rho}_{\boldsymbol{k}}\hat{\rho}_{\boldsymbol{k}^{\dagger}}}(\omega)$ of the density response function, whose respective reciprocal space Kubo formula reads as\cite{GiulianiVignale}
\begin{align}
\chi(\boldsymbol{k},\omega)=-\frac{\imath}{\hbar{V}}\lim_{\eta\to0^{+}}\int_0^{\infty}\left\langle\left[\hat{\rho}_{\boldsymbol{k}}(t),\hat{\rho}^{\dagger}(\boldsymbol{k})\right]\right\rangle_0e^{\imath(\omega+\imath\eta)t}dt\,.\label{eq:standardKubo}
\end{align}
The \emph{fluctuation--dissipation theorem} (FDT) connects the DSF with the imaginary part of the density response function according to\cite{GiulianiVignale,IchimaruBook}
\begin{align}
S(\boldsymbol{k},\omega)=-\frac{\hbar}{\pi{n}}\frac{\Im\{\chi(\boldsymbol{k},\omega)\}}{1-e^{-\beta\hbar\omega}}\,.\label{eq:quantumFDT}
\end{align}
The substitution of the FDT in the frequency integral definition of the SSF, Eq.(\ref{eq:quantumFDT}) into Eq.(\ref{eq:SSFdef}), eventually introduces the imaginary bosonic Matsubara frequencies $\imath\omega_l=2\pi\imath{l}/\beta\hbar$ and ultimately leads to a \emph{Matsubara series expansion} for the SSF that reads as\cite{DornheimPhysRep2018,TanakaJPSJ1986}
\begin{align}
S(\boldsymbol{k})=-\frac{1}{{n}\beta}\displaystyle\sum_{l=-\infty}^{\infty}\widetilde{\chi}(\boldsymbol{k},\imath\omega_l)\,.\label{eq:MatsubaraSeries}
\end{align}
The $\,\widetilde{}\,$ symbol over dynamic quantities signifies analytic continuation from the real frequency domain $\omega$ to the complex frequency domain $z$. In a similar fashion, the substitution of the FDT in the Laplace transform definition of the ITCF, Eq.(\ref{eq:quantumFDT}) into Eq.(\ref{eq:ITCFdef}), leads to a \emph{Fourier--Matsubara series expansion} for the ITCF that generalizes Eq.(\ref{eq:MatsubaraSeries}) to finite imaginary times
and reads as\cite{ToliasJCP2024,DornheimPRB2024}
\begin{align}
F(\boldsymbol{k},\tau)=-\frac{1}{n\beta}\sum_{l=-\infty}^{+\infty}\widetilde{\chi}(\boldsymbol{k},\imath\omega_l)e^{-\imath\hbar\omega_l\tau}\,.\label{eq:FourierMatsubaraSeries}
\end{align}
The \emph{detailed balance relation} is a consequence of the Lehmann representation of the DSF that is also encoded in the FDT\cite{GiulianiVignale,IchimaruBook};
\begin{align}
S(\boldsymbol{k},-\omega)=e^{-\beta\hbar\omega}S(\boldsymbol{k},\omega)\,.
\end{align}
It translates into an \emph{imaginary--time symmetry property} for the ITCF that reads as\cite{DornheimPoPRev2023,DornheimNat2022,DornheimMRE2023}
\begin{align}
F(\boldsymbol{k},\tau)=F(\boldsymbol{k},\beta-\tau)\Rightarrow{F}(\boldsymbol{k},\tau+\beta/2)=F(\boldsymbol{k},\beta/2-\tau)\,.\label{eq:ImaginaryTimeSymmetry}
\end{align}
This property implies that the imaginary-time correlation function has a minimum at $\tau=\beta/2$ regardless of the wavenumber. This explains the importance of the TSF, $S^{\beta/2}(\boldsymbol{k})=F(\boldsymbol{k},\tau=\beta/2)$, introduced above. It also explains why ITCFs are typically studied in the interval $\tau\in[0,\beta/2]$. More important, since the position of the minimum depends solely on the temperature, this symmetric expression constitutes the basis for a model-free diagnostic of the temperature directly from XRTS experiments\cite{DornheimPoPRev2023,DornheimNat2022,DornheimPoP2023,DornheimSciRep2024}. It is emphasized that all expressions introduced in this section are valid for interacting quantum many-body systems, even though they will only be utilized for the non-interacting uniform electron gas (UEG) in what follows.

\subsection{Ideal (Lindhard) density response}\label{sec:backgroundLindhard}

The density response function of the non-interacting UEG is a prerequisite for the understanding of the intricate density response of interacting electronic systems. Giuliani and Vignale offer a detailed exposition of the density response of the non-interacting UEG, still focusing mostly on the ground state\cite{GiulianiVignale}. In what follows, the subscript \enquote{$0$} will be employed to differentiate the ideal linear density response function from its interacting counterpart and the subscript \enquote{HF} will be employed to differentiate the ideal correlation functions from their interacting counterparts. After its analytic continuation in the complex frequency $z-$plane, the non-interacting (Lindhard) density response can be compactly written as\cite{GiulianiVignale}
\begin{align}
\widetilde{\chi}_0(\boldsymbol{k},z)=-\frac{2}{V}\sum_{\boldsymbol{q}}\frac{f_0\left(\boldsymbol{q}+\boldsymbol{k}\right)-f_0\left(\boldsymbol{q}\right)}{\hbar{z}-\epsilon_{\boldsymbol{q}+\boldsymbol{k}}+\epsilon_{\boldsymbol{q}}}\,,\label{eq:Lindhard0}
\end{align}
where $\epsilon_{\boldsymbol{q}}=(\hbar^2/2m)q^2$ is the single electron kinetic energy and $f_0(\boldsymbol{q})$ is the Fermi-Dirac distribution. In the thermodynamic limit, this becomes
\begin{align}
\widetilde{\chi}_0(\boldsymbol{k},z)=-2\int\frac{d^3q}{(2\pi)^3}\frac{f_0\left(\boldsymbol{q}+\boldsymbol{k}\right)-f_0\left(\boldsymbol{q}\right)}{\hbar{z}-\epsilon_{\boldsymbol{q}+\boldsymbol{k}}+\epsilon_{\boldsymbol{q}}}\,.\label{eq:Lindhard1}
\end{align}
For further postprocessing, it is more convenient to unify the arguments of the distribution function. This leads to
\begin{align}
\widetilde{\chi}_0(\boldsymbol{k},z)=-2\int\frac{d^3q}{(2\pi)^3}f_0\left(\boldsymbol{q}\right)\left[\frac{1}{\hbar{z}+\frac{\hbar^2}{2m}k^2-\frac{\hbar^2}{m}\left(\boldsymbol{q}\cdot\boldsymbol{k}\right)}-\frac{1}{\hbar{z}-\frac{\hbar^2}{2m}k^2+\frac{\hbar^2}{m}\left(\boldsymbol{q}\cdot\boldsymbol{k}\right)}\right]\,.\label{eq:Lindhard2}
\end{align}
\emph{For real frequencies}, $z=\omega+\imath0$, the application of the Sokhotski-Plemelj formula and the utilization of spherical coordinates yield the following expressions for the real and imaginary parts of the ideal Lindhard density response
\begin{align}
\Re\{\chi_0(\boldsymbol{k},\omega)\}&=\frac{2}{(2\pi)^2}\frac{m}{\hbar^2}\int_0^{\infty}\frac{q}{k}f_0(q)\ln{\left|\frac{\left(q-\frac{1}{2}k\right)^2-\frac{m^2}{\hbar^2}\frac{\omega^2}{k^2}}{\left(q+\frac{1}{2}k\right)^2-\frac{m^2}{\hbar^2}\frac{\omega^2}{k^2}}\right|}{d}q\,,\label{eq:realLindhard}\\
\Im\{\chi_0(\boldsymbol{k},\omega)\}&=\frac{1}{2\pi}\frac{m}{\hbar^2}\int_0^{\infty}\frac{q}{k}f_0(q)\left\{\mathrm{H}\left[q^2-\left(\frac{k}{2}+\frac{m}{\hbar}\frac{\omega}{k}\right)^2\right]-\mathrm{H}\left[q^2-\left(\frac{k}{2}-\frac{m}{\hbar}\frac{\omega}{k}\right)^2\right]\right\}{d}q\,.\label{eq:imagLindhard}
\end{align}
Substitution of the Fermi-Dirac distribution and introduction of the normalizations $x=k/k_{\mathrm{F}}$, $y=k/k_{\mathrm{F}}$, $\Omega=\hbar\omega/E_{\mathrm{f}}$, $\Theta=T/E_{\mathrm{f}}$, where $k_{\mathrm{F}}=(3\pi^2n)^{1/3}$ is the Fermi wavenumber and $E_{\mathrm{F}}=\epsilon_{k_{\mathrm{F}}}=(\hbar^2/2m)k_{\mathrm{F}}^2$ is the Fermi energy, yield the numerically convenient expressions\cite{GiulianiVignale}
\begin{align}
\Re\left\{\frac{\chi_0(x,\Omega)}{n\beta}\right\}&=\frac{3}{4}\frac{\Theta}{x}\int_0^{\infty}\frac{y}{\displaystyle\exp{\left(\frac{y^2}{\Theta}-\bar{\mu}\right)}+1}\ln{\left|\frac{\left(y-\frac{1}{2}x\right)^2-\left(\frac{\Omega}{2x}\right)^2}{\left(y+\frac{1}{2}x\right)^2-\left(\frac{\Omega}{2x}\right)^2}\right|}{d}y\,,\label{eq:realLindhardnorm}\\
\Im\left\{\frac{\chi_0(x,\Omega)}{n\beta}\right\}&=\frac{3\pi}{4}\frac{\Theta}{x}\int_0^{\infty}\frac{y}{\exp{\left(\frac{y^2}{\Theta}-\bar{\mu}\right)}+1}\left\{\mathrm{H}\left[y^2-\left(\frac{x}{2}+\frac{1}{2}\frac{\Omega}{x}\right)^2\right]-\mathrm{H}\left[y^2-\left(\frac{x}{2}-\frac{1}{2}\frac{\Omega}{x}\right)^2\right]\right\}{d}y\,,\label{eq:imagLindhardnorm}
\end{align}
where $\bar{\mu}=\mu/T$ with $\mu$ the chemical potential. The integration that involves the Heaviside step function can be ultimately carried out yielding a closed form expression for the imaginary part of the ideal Lindhard density response\cite{GiulianiVignale}
\begin{align}
\Im\left\{\frac{\chi_0(x,\Omega)}{n\beta}\right\}&=\frac{3\pi}{8x}\Theta^2\ln{\left\{\frac{1+\exp{\left[\bar{\mu}-\frac{1}{\Theta}\left(\frac{x}{2}+\frac{1}{2}\frac{\Omega}{x}\right)^2\right]}}{1+\exp{\left[\bar{\mu}-\frac{1}{\Theta}\left(\frac{x}{2}-\frac{1}{2}\frac{\Omega}{x}\right)^2\right]}}\right\}}\,.\label{eq:imagLindhardnormclosed}
\end{align}
Thus, combining with the FDT, see Eq.(\ref{eq:quantumFDT}), the above also implies that there exists a closed form expression for the DSF of the non-interacting UEG, which reads as
\begin{align}
S_{\mathrm{HF}}(x,\Omega)=-\frac{\hbar}{E_{\mathrm{f}}}\frac{3\Theta}{8x}\frac{1}{1-e^{-\Omega/\Theta}}\ln{\left\{\frac{1+\exp{\left[\bar{\mu}-\frac{1}{\Theta}\left(\frac{x}{2}+\frac{1}{2}\frac{\Omega}{x}\right)^2\right]}}{1+\exp{\left[\bar{\mu}-\frac{1}{\Theta}\left(\frac{x}{2}-\frac{1}{2}\frac{\Omega}{x}\right)^2\right]}}\right\}}\,.\label{eq:DSFclosed}
\end{align}
\emph{For imaginary frequencies} $z=\imath\omega_l$, important in view of Eq.(\ref{eq:MatsubaraSeries}) and Eq.(\ref{eq:FourierMatsubaraSeries}), the Lindhard density response is a real quantity that is given by\cite{DornheimPhysRep2018}
\begin{align}
\widetilde{\chi}_0(\boldsymbol{k},\imath\omega_l)=-\frac{2}{k}\frac{m}{\hbar^2}\int_0^{\infty}\frac{dq}{(2\pi)^2}\frac{q}{\exp{\left(\frac{\hbar^2q^2}{2mT}-\frac{\mu}{T}\right)}+1}\ln{\left[\frac{\left(k^2+2qk\right)^2+\left(\frac{4\pi{l}mT}{\hbar^2}\right)^2}{\left(k^2-2qk\right)^2+\left(\frac{4\pi{l}mT}{\hbar^2}\right)^2}\right]}\,,\label{eq:Lindhardimag}
\end{align}
or, in normalized units, by\cite{TanakaJPSJ1986}
\begin{align}
\frac{\widetilde{\chi}_0(x,\imath\omega_l)}{n\beta}=-\frac{3}{4}\frac{\Theta}{x}\int_0^{\infty}dy\frac{y}{\exp{\left(\frac{y^2}{\Theta}-\bar{\mu}\right)}+1}\ln{\left[\frac{\left(x^2+2xy\right)^2+\left(2\pi{l}{\Theta}\right)^2}{\left(x^2-2xy\right)^2+\left(2\pi{l}\Theta\right)^2}\right]}\,.\label{eq:Linhardimagnorm}
\end{align}

\section{Exact results}\label{sec:exact}

\subsection{Non-interacting pair correlation function}\label{sec:backgroundPCF}

For the non-interacting electron gas, the $N-$particle wavefunction is given by a Slater determinant with plane wave one-particle orbitals. This leads to the ideal $N-$particle density matrix (in the eigen-energy representation), the ideal reduced $s-$particle density matrices (by integrating out the $N-s$ particle coordinates), the ideal reduced $s-$particle densities (by considering the diagonal) and ultimately the non-interacting pair correlation function (by utilizing its connection with the two-particle density). The above procedure has been outlined by Mahan for the ground state\cite{MahanBook}. At finite temperatures and in the paramagnetic case, the same procedure within the canonical ensemble yields the following expression for the non-interacting PCF
\begin{align}
g_{\mathrm{HF}}(\boldsymbol{r})=1-\frac{1}{2}\left[\frac{2}{n}\frac{1}{V}\sum_{\boldsymbol{k}}f_0(\boldsymbol{k})e^{\imath\boldsymbol{k}\cdot\boldsymbol{r}}\right]^2\,.\label{eq:PCFideal1}
\end{align}
In the thermodynamic limit, this is equivalent to
\begin{align}
g_{\mathrm{HF}}(\boldsymbol{r})=1-\frac{2}{n^2}\left[\int\frac{d^3k}{(2\pi)^3}f_0(\boldsymbol{k})e^{\imath\boldsymbol{k}\cdot\boldsymbol{r}}\right]^2\,.\label{eq:PCFideal2}
\end{align}
In view of the isotropy, utilization of spherical coordinates yields
\begin{align}
g_{\mathrm{HF}}(r)=1-\frac{1}{2}\left[\frac{1}{\pi^2nr}\int_0^{\infty}k\sin{(kr)}f_0(k)dk\right]^2\,.\label{eq:PCFideal3}
\end{align}
After introduction of the normalized wavevector $y=k/k_{\mathrm{F}}$ and the normalized distance $x=rk_{\mathrm{F}}$ as well as substitution of the Fermi-Dirac distribution, one obtains the non-interacting PCF expression
\begin{align}
g_{\mathrm{HF}}(x)=1-\frac{1}{2}\frac{9}{x^2}\left[\int_0^{\infty}\frac{y\sin{(xy)}}{\exp{\left(\frac{y^2}{\Theta}-\bar{\mu}\right)}+1}dy\right]^2\,.\label{eq:PCFideal}
\end{align}
The known closed form expression for the non-interacting ground state PCF directly follows, since $1/\left[\exp{\left(\frac{y^2}{\Theta}-\bar{\mu}\right)}+1\right]=\mathrm{H}(1-y)$ as $\Theta\to0$. After introducing the spherical Bessel function of the first kind and first order, $\mathrm{j}_1(x)=(\sin{x}-x\cos{x})/x^2$, it reads as\cite{MahanBook}
\begin{align}
g_{\mathrm{HF}}(x)=1-\frac{1}{2}\frac{9}{x^2}\left[\frac{\sin{x}-x\cos{x}}{x^2}\right]^2=1-\frac{1}{2}\frac{9}{x^2}\mathrm{j}_1^2(x)\,.\label{eq:PCFidealground}
\end{align}
Finite temperature corrections to the ground state result can be obtained by employing the Sommerfeld expansion and keeping up to the first order term. This leads to the $\Theta\ll1$ expression
\begin{align}
g_{\mathrm{HF}}(x)\simeq1-\frac{1}{2}\frac{9}{x^2}\left[\frac{\sin{x}}{x^2}-\frac{\cos{x}}{x}+\frac{\pi^2}{24}\Theta^2x\cos{x}\right]^2\,.\label{eq:PCFideallow}
\end{align}
On the other hand, the high temperature limit can be obtained by substituting the Fermi-Dirac distribution with the Maxwellian distribution. Ultimately, after disposing of the normalized chemical potential through the normalization condition, this leads to the $\Theta\gg1$ expression
\begin{align}
g_{\mathrm{HF}}(x)\simeq1-\frac{1}{2}\exp{\left(-\frac{\Theta}{2}x^2\right)}\,.\label{eq:PCFidealhigh}
\end{align}
The ground state expression, Eq.(\ref{eq:PCFidealground}), remains very accurate up to $\Theta\simeq0.1$ with errors emerging at the intermediate distance range, $1\leq{r}k_{\mathrm{F}}\leq3$. The low temperature expression, Eq.(\ref{eq:PCFideallow}), is very accurate also up to $\Theta\simeq0.1$, but with errors now emerging at the short distance range, $rk_{\mathrm{F}}\leq2$. The high temperature expression, Eq.(\ref{eq:PCFidealhigh}), is near exact even at $\Theta\simeq2$, with small errors emerging from $\Theta\simeq1$ at the intermediate distance range, $1\leq{r}k_{\mathrm{F}}\leq3$. It is worth pointing out that the finite temperature and ground state non-interacting PCF expressions naturally comply with the Stillinger-Lovett sum rule\cite{BausPhysRep1980}, $n\int[g(r)-1]d^3r=-1$, that expresses overall charge neutrality. On the other hand, the approximate low temperature and high temperature PCF expressions both violate the Stillinger-Lovett sum rule.

\subsection{Non-interacting static structure factor}\label{sec:SSFnew}

\emph{A numerically convenient expression for the non-interacting SSF} can be obtained by substituting the Lindhard response function evaluated at the imaginary bosonic Matsubara frequencies, Eq.(\ref{eq:Linhardimagnorm}), into the Matsubara series expansion for the SSF, Eq.(\ref{eq:MatsubaraSeries});
\begin{align*}
S_{\mathrm{HF}}(x)=\frac{3\Theta}{4x}\displaystyle\sum_{l=-\infty}^{\infty}\int_0^{\infty}dy\frac{y}{\exp{\left(\frac{y^2}{\Theta}-\bar{\mu}\right)}+1}\ln{\left[\frac{\left(x^2+2xy\right)^2+\left(2\pi{l}{\Theta}\right)^2}{\left(x^2-2xy\right)^2+\left(2\pi{l}\Theta\right)^2}\right]}\,.\nonumber
\end{align*}
Interchanging the series and integral operators, after utilizing the basic logarithmic property, one obtains
\begin{align*}
S_{\mathrm{HF}}(x)=\frac{3\Theta}{4x}\int_0^{\infty}dy\frac{y}{\exp{\left(\frac{y^2}{\Theta}-\bar{\mu}\right)}+1}\ln{\left\{\prod_{l=-\infty}^{\infty}\left[\frac{\left(\frac{x^2+2xy}{2\Theta}\right)^2+\left(\pi{l}\right)^2}{\left(\frac{x^2-2xy}{2\Theta}\right)^2+\left(\pi{l}\right)^2}\right]\right\}}\,.\nonumber
\end{align*}
Employing the identity $\prod_{n=-\infty}^{+\infty}\left(a^2+\pi^2n^2\right)/\left(b^2+\pi^2n^2\right)=\sinh^2{(a)}/\sinh^2{(b)}$, rearranging the exponential terms and utilizing the normalization condition, one ends up with
\begin{align*}
S_{\mathrm{HF}}(x)=1+\frac{3\Theta}{2x}\int_0^{\infty}dy\frac{y}{\exp{\left(\frac{y^2}{\Theta}-\bar{\mu}\right)}+1}\ln{\left\{\left|\frac{1-\exp{\left[-\frac{(x^2+2xy)}{\Theta}\right]}}{1-\exp{\left[-\frac{(x^2-2xy)}{\Theta}\right]}}\right|\right\}}\,.\nonumber
\end{align*}
The same expression for the non-interacting SSF can be obtained by substituting the integral form of the imaginary part of the Lindhard response function, Eq.(\ref{eq:imagLindhardnorm}), into the FDT, Eq.(\ref{eq:quantumFDT}) and then into the zero frequency moment sum rule that defines the SSF, Eq.(\ref{eq:SSFdef});
\begin{align*}
S_{\mathrm{HF}}(x)=-\frac{3}{4}\frac{1}{x}\int_0^{\infty}\frac{y}{\exp{\left(\frac{y^2}{\Theta}-\bar{\mu}\right)}+1}\int_{-\infty}^{+\infty}\frac{1}{1-e^{-\Omega/\Theta}}\left\{\mathrm{H}\left[y^2-\left(\frac{x}{2}+\frac{1}{2}\frac{\Omega}{x}\right)^2\right]-\mathrm{H}\left[y^2-\left(\frac{x}{2}-\frac{1}{2}\frac{\Omega}{x}\right)^2\right]\right\}d\Omega{d}y\,.
\end{align*}
The frequency integral is considered first. After some re-arrangements, a change of variables and some exponential algebra, one has
\begin{align*}
S_{\mathrm{HF}}(x)=-\frac{3}{4}\frac{1}{x}\int_0^{\infty}\frac{y}{\exp{\left(\frac{y^2}{\Theta}-\bar{\mu}\right)}+1}\left\{\int_{-\infty}^{+\infty}\coth{\left(\frac{\Omega}{2\Theta}\right)}\mathrm{H}\left[y^2-\left(\frac{x}{2}+\frac{1}{2}\frac{\Omega}{x}\right)^2\right]d\Omega\right\}{d}y\,.\nonumber
\end{align*}
The step function can be removed, provided that the infinite integration boundary is adjusted accordingly. The change of variables $z=\Omega/(2\Theta)$ is carried out and the integration is performed, courtesy of $\int\coth{z}dz=\ln{|\sinh{z}|}$, leading to
\begin{align*}
S_{\mathrm{HF}}(x)=+\frac{3\Theta}{2x}\int_0^{\infty}\frac{y}{\exp{\left(\frac{y^2}{\Theta}-\bar{\mu}\right)}+1}\ln{\left[\left|\sinh{\left(\frac{x^2+2xy}{2\Theta}\right)}/\sinh{\left(\frac{x^2-2xy}{2\Theta}\right)}\right|\right]}{d}y\,.
\end{align*}
Rearranging the exponential terms and utilizing the normalization condition, one again obtains
\begin{align}
S_{\mathrm{HF}}(x)=1+\frac{3\Theta}{2x}\int_0^{\infty}dy\frac{y}{\exp{\left(\frac{y^2}{\Theta}-\bar{\mu}\right)}+1}\ln{\left\{\left|\frac{1-\exp{\left[-\frac{(x^2+2xy)}{\Theta}\right]}}{1-\exp{\left[-\frac{(x^2-2xy)}{\Theta}\right]}}\right|\right\}}\,.\label{idealSSFform1}
\end{align}
\emph{An equivalent expression for the non-interacting SSF} can be obtained by substituting the closed form expression for the non-interacting DSF, Eq.(\ref{eq:DSFclosed}), into the zero frequency moment sum rule that defines the SSF, Eq.(\ref{eq:SSFdef});
\begin{align*}
S_{\mathrm{HF}}(x)=-\frac{3\Theta}{8x}\int_{-\infty}^{+\infty}\frac{1}{1-e^{-\Omega/\Theta}}\ln{\left\{\frac{1+\exp{\left[\bar{\mu}-\frac{1}{\Theta}\left(\frac{x}{2}+\frac{1}{2}\frac{\Omega}{x}\right)^2\right]}}{1+\exp{\left[\bar{\mu}-\frac{1}{\Theta}\left(\frac{x}{2}-\frac{1}{2}\frac{\Omega}{x}\right)^2\right]}}\right\}}d\Omega\,.\nonumber
\end{align*}
After some re-arrangements, a change of variables and some exponential algebra, the hyperbolic cotangent emerges in
\begin{align*}
S_{\mathrm{HF}}(x)=+\frac{3\Theta}{8x}\int_{0}^{+\infty}\coth{\left(\frac{\Omega}{2\Theta}\right)}\ln{\left\{\frac{1+\exp{\left[\bar{\mu}-\frac{1}{\Theta}\left(\frac{x}{2}-\frac{1}{2}\frac{\Omega}{x}\right)^2\right]}}{1+\exp{\left[\bar{\mu}-\frac{1}{\Theta}\left(\frac{x}{2}+\frac{1}{2}\frac{\Omega}{x}\right)^2\right]}}\right\}}d\Omega\,.\nonumber
\end{align*}
Ultimately, the change of variables $\Omega=xy$ leads to
\begin{align}
S_{\mathrm{HF}}(x)=+\frac{3\Theta}{8}\int_{0}^{+\infty}\coth{\left(\frac{xy}{2\Theta}\right)}\ln{\left\{\frac{1+\exp{\left[\bar{\mu}-\frac{(x-y)^2}{4\Theta}\right]}}{1+\exp{\left[\bar{\mu}-\frac{(x+y)^2}{4\Theta}\right]}}\right\}}dy\,.\label{idealSSFform2}
\end{align}
\emph{Another equivalent expression for the non-interacting SSF} can be obtained by rearranging the non-interacting PCF, Eq.(\ref{eq:PCFideal}). After exploiting the dummy nature of the integration variables in the square of the integral, one obtains
\begin{align*}
g_{\mathrm{HF}}(x)=1-\frac{1}{2}\frac{9}{x^2}\int_0^{\infty}\int_0^{\infty}zy\sin{(xy)}\sin{(xz)}\frac{1}{\exp{\left(\frac{y^2}{\Theta}-\bar{\mu}\right)}+1}\frac{1}{\exp{\left(\frac{z^2}{\Theta}-\bar{\mu}\right)}+1}dydz\,.\nonumber
\end{align*}
The well-known trigonometric product formula $\sin{(a)}\sin{(b)}=(1/2)[\cos{(a-b)}-\cos{(a+b)}]$ is employed. The double integral within the integration domain $(0,+\infty)\times(0,+\infty)$ is split into two double integrals. After a change of variables, the double integrals are unified into a considerably simpler double integral, but within the $(0,+\infty)\times(-\infty,+\infty)$ integration domain.
\begin{align*}
g_{\mathrm{HF}}(x)=1-\frac{1}{4}\frac{9}{x^2}\int_0^{\infty}\int_{-\infty}^{+\infty}\frac{u}{\exp{\left(\frac{u^2}{\Theta}-\bar{\mu}\right)}+1}\frac{(u-w)\cos{(wx)}}{\exp{\left[\frac{(u-w)^2}{\Theta}-\bar{\mu}\right]}+1}dudw\,.\nonumber
\end{align*}
The coordinate transformation ($y=u, z=u-w$), that has a unity Jacobian determinant, is employed.  This is followed by integration by parts, concerning the $w-$dependent integral. This yields a more compact double integral within the $(0,+\infty)\times(-\infty,+\infty)$ integration domain.
\begin{align*}
g_{\mathrm{HF}}(x)=1-\frac{9\Theta}{8x}\int_0^{\infty}\frac{u}{\exp{\left(\frac{u^2}{\Theta}-\bar{\mu}\right)}+1}\left\{\int_{-\infty}^{+\infty}\sin{(wx)}\ln{\left|1+\exp{\left[\bar{\mu}-\frac{(u-w)^2}{\Theta}\right]}\right|}dw\right\}du\,.\nonumber
\end{align*}
After some re-arrangements and another change of variables, the integration domain is shrunk to $(0,+\infty)\times(0,+\infty)$. Some logarithmic algebra and the introduction of the TCF lead to
\begin{align*}
h_{\mathrm{HF}}(x)=\frac{3}{2x}\int_0^{\infty}y\sin{(xy)}\left\{1-\frac{3\Theta}{4y}\int_{0}^{+\infty}\frac{z}{\exp{\left(\frac{z^2}{\Theta}-\bar{\mu}\right)}+1}\ln{\left|\frac{1+\exp{\left[\bar{\mu}-\frac{(z-y)^2}{\Theta}\right]}}{1+\exp{\left[\bar{\mu}-\frac{(z+y)^2}{\Theta}\right]}}\right|}dz-1\right\}dy\,.\nonumber
\end{align*}
The spatial Fourier transform connection between the SSF and the TCF, Eq.(\ref{eq:microconnection}) becomes
\begin{align*}
h_{\mathrm{HF}}(x)=\frac{3}{2x}\int_0^{\infty}y\sin{(xy)}\left\{S_{\mathrm{HF}}(y)-1\right\}dy\,.\nonumber
\end{align*}
for isotropic systems. In view of the Fourier uniqueness theorem, a direct comparison between the last two expressions ultimately yields
\begin{align}
S_{\mathrm{HF}}(x)=1-\frac{3\Theta}{4x}\int_{0}^{+\infty}\frac{y}{\exp{\left(\frac{y^2}{\Theta}-\bar{\mu}\right)}+1}\ln{\left|\frac{1+\exp{\left[\bar{\mu}-\frac{(y-x)^2}{\Theta}\right]}}{1+\exp{\left[\bar{\mu}-\frac{(y+x)^2}{\Theta}\right]}}\right|}dy\,.\label{idealSSFform3}
\end{align}
The equivalence of the three non-interacting SSF expressions, Eqs.(\ref{idealSSFform1},\ref{idealSSFform2},\ref{idealSSFform3}) has been confirmed numerically. All three numerical quadratures are rapidly converging with standard numerical techniques. The correct \emph{ground state result} can also be confirmed, e.g. starting from Eq.(\ref{idealSSFform3}). Courtesy of the limit
\begin{align*}
\displaystyle\lim_{\Theta\to0}\Theta\ln{\left|1+\exp{\left[\bar{\mu}-\frac{(y-x)^2}{\Theta}\right]}\right|}=\left[1-(y-x)^2\right]\mathrm{H}\left[1-(y-x)^2\right]\,,\nonumber
\end{align*}
one directly obtains
\begin{align*}
S_{\mathrm{HF}}(x)=1-\frac{3}{4x}\int_{-\infty}^{+\infty}y\left[1-(y-x)^2\right]\mathrm{H}(1-y)\mathrm{H}\left[1-(y-x)^2\right]dy\,.\nonumber
\end{align*}
For $x>2$, this leads to
\begin{align*}
S_{\mathrm{HF}}(x>2)=1\,.\nonumber
\end{align*}
For $x<2$, after some straightforward calculations, this leads to
\begin{align*}
S_{\mathrm{HF}}(x<2)&=\frac{1}{16}x\left(12-x^2\right)\nonumber\,.
\end{align*}
Combining the above, one indeed obtains the ground state non-interacting static structure factor\cite{MahanBook}
\begin{align}
S_{\mathrm{HF}}(x)=1+\left[\frac{x}{16}(12-x^2)-1\right]\mathrm{H}(2-x)\,.\label{idealSSFground}
\end{align}
For completeness, the \emph{high temperature limit} will also be considered, e.g. starting from Eq.(\ref{idealSSFform3}). The standard substitution of the Fermi Dirac distribution with the Maxwellian distribution yields
\begin{align*}
S_{\mathrm{HF}}(x)\simeq1-\frac{3\Theta}{4x}e^{2\bar{\mu}}\int_0^{\infty}y\exp{\left(-\frac{y^2}{\Theta}\right)}\left\{\exp{\left[-\frac{(y-x)^2}{\Theta}\right]}-\exp{\left[-\frac{(y+x)^2}{\Theta}\right]}\right\}dy\,.\nonumber
\end{align*}
The above expression can be conveniently recast into
\begin{align*}
S_{\mathrm{HF}}(x)\simeq1-\frac{3\Theta}{4x}e^{2\bar{\mu}}\exp{\left(-\frac{x^2}{\Theta}\right)}\int_{-\infty}^{\infty}y\exp{\left[-\frac{2y^2-2xy}{\Theta}\right]}dy\,.\nonumber
\end{align*}
The integral is of the Gaussian integral form, $\int_{-\infty}^{\infty}y\exp{[-(2y^2-2xy)/\Theta]}dy=(1/2)\sqrt{\pi/2}\sqrt{\Theta}x\exp{[x^2/(2\Theta)]}$. Moreover, in the high temperature limit $\Theta\gg1$, the normalized chemical potential obeys $(3/4)\sqrt{\pi}\Theta^{3/2}e^{\bar{\mu}}=1$ which can be solved for the thermodynamic activity $\lambda=e^{\bar{\mu}}$ to yield $\lambda=(4/3)(1/\sqrt{\pi})\Theta^{-3/2}$. Combining the above, one ends up with
\begin{align}
S_{\mathrm{HF}}(x)&\simeq1-\sqrt{\frac{2}{\pi\Theta}}\frac{1}{3\Theta}\exp{\left(-\frac{x^2}{2\Theta}\right)}\,.\label{idealSSFhighT}
\end{align}
It is noted that the same result emerges when combining the high temperature limit of the non-interacting PCF, Eq.(\ref{eq:PCFidealhigh}), with the Fourier transform connection between the SSF and the TCF, Eq.(\ref{eq:microconnection}). The ground state expression, Eq.(\ref{idealSSFground}), remains very accurate up to $\Theta\simeq0.02$ with errors emerging at the long wavelength range ${k}/k_{\mathrm{F}}\leq0.2$. The high temperature expression, Eq.(\ref{idealSSFhighT}), is near exact even at $\Theta\simeq2$, with small errors emerging from $\Theta\simeq1$ at the short-to-intermediate wavenumber range, ${k}/k_{\mathrm{F}}\leq1.5$.

\subsection{Non-interacting imaginary--time correlation function}\label{sec:ITCFnew}

\emph{A numerically convenient expression for the non-interacting ITCF} can be obtained by substituting the Lindhard response function evaluated at the imaginary bosonic Matsubara frequencies, Eq.(\ref{eq:Linhardimagnorm}), into the Fourier--Matsubara series expansion for the ITCF, Eq.(\ref{eq:FourierMatsubaraSeries}). Introducing the imaginary time normalization $\tau^{\star}=T\tau$ that guarantees a state independent imaginary time interval and interchanging the series and integral operators, one has
\begin{align*}
F_{\mathrm{HF}}(x,\tau^{\star})&=\frac{3}{4}\frac{\Theta}{x}\int_0^{\infty}dy\frac{y}{\exp{\left(\displaystyle\frac{y^2}{\Theta}-\bar{\mu}\right)}+1}\left\{\sum_{l=-\infty}^{+\infty}\ln{\left[\frac{\left(\displaystyle\frac{x^2+2xy}{2\pi\Theta}\right)^2+{l}^2}{\left(\displaystyle\frac{x^2-2xy}{2\pi\Theta}\right)^2+l^2}\right]}e^{2\pi\imath{\tau^{\star}}{l}}\right\}\,.\nonumber
\end{align*}
For the evaluation of the infinite series, an interlude to a special function known as Hurwitz-Lerch zeta function is essential. The Hurwitz-Lerch zeta function is straighforwardly connected to the Lerch transcendent, which is a generalization of the Hurwitz zeta function (itself a generalization of the Riemann zeta function). The Lerch trancendent and the Hurwitz-Lerch zeta function are defined by
\begin{align*}
\Phi(z,s,a)&=\sum_{n=0}^{\infty}\frac{z^n}{(n+a)^s}\,,\nonumber\\
L(\lambda,s,a)&=\sum_{n=0}^{\infty}\frac{e^{2\pi\imath\lambda{n}}}{(n+a)^s}\,,\nonumber
\end{align*}
respectively. The connection $L(\lambda,s,a)=\Phi(e^{2\pi\imath\lambda},s,a)$ is apparent. Differentiation of $\Phi(z,s,a)$ with respect to the $s-$argument and consideration of the limit $s\to0$ leads to
\begin{align*}
\left.\frac{\partial\Phi(z,s,a)}{\partial{s}}\right|_{s=0}&=-\sum_{n=0}^{\infty}z^n\ln{(n+a)}\,,\nonumber
\end{align*}
which can be utilized to derive the identity
\begin{align*}
\sum_{n=0}^{+\infty}\ln{\left(\frac{a^2+n^2}{b^2+n^2}\right)}e^{\pm2\pi\imath{c}n}&=-\left.\frac{\partial\Phi\left(e^{\pm2\pi\imath{c}},s,\imath{a}\right)}{\partial{s}}\right|_{s=0}-\left.\frac{\partial\Phi\left(e^{\pm2\pi\imath{c}},s,-\imath{a}\right)}{\partial{s}}\right|_{s=0}+\left.\frac{\partial\Phi\left(e^{\pm2\pi\imath{c}},s,\imath{b}\right)}{\partial{s}}\right|_{s=0}+\left.\frac{\partial\Phi\left(e^{\pm2\pi\imath{c}},s,-\imath{b}\right)}{\partial{s}}\right|_{s=0}\,,\nonumber
\end{align*}
which can be utilized to derive the identity
\begin{align*}
\sum_{n=-\infty}^{+\infty}\ln{\left(\frac{a^2+n^2}{b^2+n^2}\right)}e^{2\pi\imath{c}n}&=-2\ln{\left|\frac{a}{b}\right|}-\left.\frac{\partial{L}\left(-c,s,\imath{a}\right)}{\partial{s}}\right|_{s=0}-\left.\frac{\partial{L}\left(-c,s,-\imath{a}\right)}{\partial{s}}\right|_{s=0}+\left.\frac{\partial{L}\left(-c,s,\imath{b}\right)}{\partial{s}}\right|_{s=0}+\left.\frac{\partial{L}\left(-c,s,-\imath{b}\right)}{\partial{s}}\right|_{s=0}\nonumber\\&\quad-\left.\frac{\partial{L}\left(+c,s,\imath{a}\right)}{\partial{s}}\right|_{s=0}-\left.\frac{\partial{L}\left(+c,s,-\imath{a}\right)}{\partial{s}}\right|_{s=0}+\left.\frac{\partial{L}\left(+c,s,\imath{b}\right)}{\partial{s}}\right|_{s=0}+\left.\frac{\partial{L}\left(+c,s,-\imath{b}\right)}{\partial{s}}\right|_{s=0}\,.\nonumber
\end{align*}
The latter identity directly leads to
\begin{align}
F_{\mathrm{HF}}(x,\tau^{\star})&=\frac{3}{4}\frac{\Theta}{x}\int_0^{\infty}dy\frac{y}{\exp{\left(\displaystyle\frac{y^2}{\Theta}-\bar{\mu}\right)}+1}\left\{-2\ln{\left|\frac{x+2y}{x-2y}\right|}-\left.\frac{\partial{L}\left(-{\tau^{\star}},s,\imath\displaystyle\frac{x^2+2xy}{2\pi\Theta}\right)}{\partial{s}}\right|_{s=0}-\left.\frac{\partial{L}\left(-{\tau^{\star}},s,-\imath\displaystyle\frac{x^2+2xy}{2\pi\Theta}\right)}{\partial{s}}\right|_{s=0}\right.\nonumber\\&\quad\left.+\left.\frac{\partial{L}\left(-{\tau^{\star}},s,\imath\displaystyle\frac{x^2-2xy}{2\pi\Theta}\right)}{\partial{s}}\right|_{s=0}+\left.\frac{\partial{L}\left(-{\tau^{\star}},s,-\imath\displaystyle\frac{x^2-2xy}{2\pi\Theta}\right)}{\partial{s}}\right|_{s=0}-\left.\frac{\partial{L}\left(\tau^{\star},s,\imath\displaystyle\frac{x^2+2xy}{2\pi\Theta}\right)}{\partial{s}}\right|_{s=0}\right.\nonumber\\&\quad\left.-\left.\frac{\partial{L}\left(\tau^{\star},s,-\imath\displaystyle\frac{x^2+2xy}{2\pi\Theta}\right)}{\partial{s}}\right|_{s=0}+\left.\frac{\partial{L}\left(\tau^{\star},s,\imath\displaystyle\frac{x^2-2xy}{2\pi\Theta}\right)}{\partial{s}}\right|_{s=0}+\left.\frac{\partial{L}\left(\tau^{\star},s,-\imath\displaystyle\frac{x^2-2xy}{2\pi\Theta}\right)}{\partial{s}}\right|_{s=0}\right\}\,.\label{idealITCFform1}
\end{align}
\emph{An equivalent expression for the non-interacting ITCF} can be obtained by substituting the closed form expression for the non-interacting DSF, Eq.(\ref{eq:DSFclosed}), into the two-sided Laplace transform that defines the ITCF, Eq.(\ref{eq:ITCFdef});
\begin{align*}
F_{\mathrm{HF}}(x,\tau^{\star})&=-\frac{3\Theta}{8x}\int_{-\infty}^{+\infty}\frac{e^{-(\Omega/\Theta)\tau^{\star}}}{1-e^{-\Omega/\Theta}}\ln{\left\{\frac{1+\exp{\left[\bar{\mu}-\frac{1}{\Theta}\left(\frac{x}{2}+\frac{1}{2}\frac{\Omega}{x}\right)^2\right]}}{1+\exp{\left[\bar{\mu}-\frac{1}{\Theta}\left(\frac{x}{2}-\frac{1}{2}\frac{\Omega}{x}\right)^2\right]}}\right\}}d\Omega\,.\nonumber
\end{align*}
Decomposition of the $(-\infty,+\infty)$ integration interval into the positive $(0,+\infty)$ and negative $(-\infty,0)$ interval, introduction of the change of variables $\Omega\to-\Omega$ in the negative interval integral, unification of the integrals into a single positive frequency integral, some cumbersome exponential algebra and introduction of hyperbolic trigonometric functions, yield
\begin{align*}
F_{\mathrm{HF}}(x,\tau^{\star})&=-\frac{3\Theta}{8x}\int_{0}^{+\infty}\frac{\cosh{\left[\frac{\Omega}{\Theta}(\tau^{\star}-1/2)\right]}}{\sinh{\left(\frac{\Omega}{2\Theta}\right)}}\ln{\left\{\frac{1+\exp{\left[\bar{\mu}-\frac{1}{\Theta}\left(\frac{x}{2}+\frac{1}{2}\frac{\Omega}{x}\right)^2\right]}}{1+\exp{\left[\bar{\mu}-\frac{1}{\Theta}\left(\frac{x}{2}-\frac{1}{2}\frac{\Omega}{x}\right)^2\right]}}\right\}}d\Omega\,.\nonumber
\end{align*}
After the change of variables $\Omega\to{xy}$, which allows a significant simplification of the argument of the natural logarithm, one ends up with
\begin{align}
F_{\mathrm{HF}}(x,\tau^{\star})&=+\frac{3\Theta}{8}\int_{0}^{+\infty}\frac{\cosh{\left[\frac{xy}{\Theta}(\tau^{\star}-1/2)\right]}}{\sinh{\left(\frac{xy}{2\Theta}\right)}}\ln{\left\{\frac{1+\exp{\left[\bar{\mu}-\frac{(x-y)^2}{4\Theta}\right]}}{1+\exp{\left[\bar{\mu}-\frac{(x+y)^2}{4\Theta}\right]}}\right\}}dy\,.\label{idealITCFform2}
\end{align}
It is evident that for $\tau^{\star}=0$, Eq.(\ref{idealITCFform2}) becomes identical to Eqs.(\ref{idealSSFform2}), as expected given $S_{\mathrm{HF}}(x)=F_{\mathrm{HF}}(x,0)$. The equivalence of the two non-interacting ITCF expressions, Eqs.(\ref{idealITCFform1},\ref{idealITCFform2}) has been confirmed numerically. However, Eq.(\ref{idealITCFform2}) is much more convenient for numerical evaluations even when employing software that include tabulations of the Hurwitz-Lerch zeta function. For completeness, we note the short and long-wavelength limits of the non-interacting ITCF. After successive Taylor expansions, it can be shown that the long wavelength limit does not depend on the imaginary time and is given by
\begin{align}
F_{\mathrm{HF}}(x\to0,\tau^{\star})&=+\frac{3}{2}\Theta\int_{0}^{+\infty}\frac{1}{\exp{\left(\frac{y^2}{\Theta}-\bar{\mu}\right)}+1}dy\,.\label{idealITCFlong}
\end{align}
After successive asymptotic expansions, it can also be shown that the short wavelength limit depends on the imaginary time in a discontinuous manner and is given by
\begin{align}
F_{\mathrm{HF}}(x\to\infty,\tau^{\star})=\delta_{\tau^{\star}0}+\delta_{\tau^{\star}1}\Rightarrow{F}_{\mathrm{HF}}(x\to\infty,\tau^{\star}\neq0,1)=0\,,{F}_{\mathrm{HF}}(x\to\infty,\tau^{\star}=0,1)&=1\,.\label{idealITCFshort}
\end{align}

\subsection{Non-interacting thermal structure factor}\label{sec:TSFnew}

\emph{A numerically convenient expression for the non-interacting TSF} can be obtained by substituting the Lindhard response function evaluated at the imaginary bosonic Matsubara frequencies, Eq.(\ref{eq:Linhardimagnorm}), into the Fourier--Matsubara series expansion for the ITCF, Eq.(\ref{eq:FourierMatsubaraSeries}) and setting $\tau^{\star}=1/2$ within the $\tau^{\star}=T\tau$ normalization. Interchanging the series and integral operators, one has
\begin{align*}
S_{\mathrm{HF}}^{\beta/2}(x)=\frac{3}{4}\frac{\Theta}{x}\int_0^{\infty}dy\frac{y}{\exp{\left(\frac{y^2}{\Theta}-\bar{\mu}\right)}+1}\sum_{l=-\infty}^{+\infty}(-1)^l\ln{\left[\frac{\left(\frac{x^2+2xy}{2\Theta}\right)^2+\pi^2{l}^2}{\left(\frac{x^2-2xy}{2\Theta}\right)^2+\pi^2{l}^2}\right]}\,.\nonumber
\end{align*}
It is preferable to avoid introducing the Hurwitz-Lerch zeta function. An alternative evaluation of the alternating infinite series is based on the infinite product identity $\prod_{n=-\infty}^{+\infty}\left(a^2+\pi^2n^2\right)/\left(b^2+\pi^2n^2\right)=\sinh^2{(a)}/\sinh^2{(b)}$. First, from the basic logarithmic property, one directly has
\begin{align*}
\displaystyle\sum_{n=-\infty}^{+\infty}\ln{\left(\frac{a^2+\pi^2n^2}{b^2+\pi^2n^2}\right)}=\ln{\left(\frac{\sinh{a}}{\sinh{b}}\right)^2}\,.\nonumber
\end{align*}
Second, separating into negative, zero and positive order terms together with parity considerations, one obtains
\begin{align*}
\displaystyle\sum_{n=0}^{+\infty}\ln{\left(\frac{a^2+\pi^2n^2}{b^2+\pi^2n^2}\right)}&=\ln{\left|\frac{a\sinh{a}}{b\sinh{b}}\right|}\,.\nonumber
\end{align*}
Third, separating into even and odd order terms, one gets
\begin{align*}
\displaystyle\sum_{n=0}^{+\infty}(-1)^n\ln{\left(\frac{a^2+\pi^2n^2}{b^2+\pi^2n^2}\right)}&=\ln{\left|\frac{a\tanh{(a/2)}}{b\tanh{(b/2)}}\right|}\,.\nonumber
\end{align*}
Finally, separating into negative, zero and positive order terms together with parity considerations, one ends up with
\begin{align*}
\displaystyle\sum_{n=-\infty}^{+\infty}(-1)^n\ln{\left(\frac{a^2+\pi^2n^2}{b^2+\pi^2n^2}\right)}&=\ln{\left[\frac{\tanh{(a/2)}}{\tanh{(b/2)}}\right]^2}\,.\nonumber
\end{align*}
The latter identity directly yields
\begin{align*}
S_{\mathrm{HF}}^{\beta/2}(x)=\frac{3}{4}\frac{\Theta}{x}\int_0^{\infty}dy\frac{y}{\exp{\left(\frac{y^2}{\Theta}-\bar{\mu}\right)}+1}\ln{\left[\frac{\tanh{\left(\frac{x^2+2xy}{4\Theta}\right)}}{\tanh{\left(\frac{x^2-2xy}{4\Theta}\right)}}\right]^2}\,.\nonumber
\end{align*}
After some elementary hyperbolic trigonometric algebra in the integrand, decomposition of the integral into two adders on the basis of the basic logarithmic property, expansion of the integration range by exploiting the parity properties of the integrand and unification of the integrals into a single $(-\infty,+\infty)$ integration interval integral, one ends up with the compact expression
\begin{align}
S_{\mathrm{HF}}^{\beta/2}(x)&=\frac{3}{2}\frac{\Theta}{x}\int_{-\infty}^{\infty}dy\frac{y}{\exp{\left(\frac{y^2}{\Theta}-\bar{\mu}\right)}+1}\ln{\left|\frac{1-\exp{\left(-\frac{x^2+2xy}{2\Theta}\right)}}{1+\exp{\left(-\frac{x^2+2xy}{2\Theta}\right)}}\right|}\,.\label{idealTSFform1}
\end{align}
\emph{An equivalent expression for the non-interacting TSF} can be obtained by substituting the closed form expression for the non-interacting DSF, Eq.(\ref{eq:DSFclosed}), into the two-sided Laplace transform that defines the ITCF, Eq.(\ref{eq:ITCFdef}) and setting $\tau^{\star}=1/2$ within the $\tau^{\star}=T\tau$ normalization. This procedure is equivalent to setting $\tau^{\star}=1/2$ in Eq.(\ref{idealITCFform2}). This leads to
\begin{align}
S_{\mathrm{HF}}^{\beta/2}(x)&=+\frac{3\Theta}{8}\int_{0}^{+\infty}\mathrm{csch}{\left(\frac{xy}{2\Theta}\right)}\ln{\left\{\frac{1+\exp{\left[\bar{\mu}-\frac{(x-y)^2}{4\Theta}\right]}}{1+\exp{\left[\bar{\mu}-\frac{(x+y)^2}{4\Theta}\right]}}\right\}}dy\,.\label{idealTSFform2}
\end{align}
The equivalence of the two non-interacting TSF expressions, Eqs.(\ref{idealTSFform1},\ref{idealTSFform2}) has also been confirmed numerically. Both numerical quadratures are rapidly converging with standard numerical techniques.

\section{Applications}\label{sec:applications}

\subsection{Non-interacting pair correlation function and the classical mapping method}\label{sec:mapping}

The classical mapping method was introduced more than twenty years ago by Perrot \& Dharma-wardana\cite{DharmawardanaPRL2000,PerrotPRB2000,DharmawardanaIJQC2012}. The basic idea is to map the fully quantum electronic system to a fictitious two electron-component classical system (see the spin-up and -down electrons) interacting with an effective pair potential, whose effective classical temperature is allowed to differ from the thermodynamic temperature. Such a quantum--to--classical mapping opens up the way for the use of the integral equation theory of liquids\cite{HansenBook} to determine the static structure and thermodynamics of the interacting UEG. In particular, the Ornstein--Zernike equation within the hypernetted--chain (HNC) closure can then be utilized to determine the PCF without any additional input\cite{HansenBook}. The original formalism is based on an empirical correspondence between the effective classical and thermodynamic temperatures that is obtained by reproducing highly accurate quantum Monte Carlo results for the ground state exchange-correlation energy. The effective potential is constructed by adding a diffraction correction to the Coulomb potential to account for delocalization within the de Broglie wavelength (Deutsch regularization) and by superimposing a Fermi--hole potential to treat Pauli exclusion effects in the non-interacting limit\cite{DharmawardanaPRL2000,PerrotPRB2000}. In order to compute the Fermi--hole potential, one utilizes the (renormalized) spin-resolved version of the non-interacting PCF, introduced in Eq.(\ref{eq:PCFideal2}), which reads as
\begin{align}
g_{\mathrm{HF}}^{\mathrm{uu}}(x)=1-\frac{9}{x^2}\left[\int_0^{\infty}\frac{y\sin{(xy)}}{\exp{\left(\frac{y^2}{\Theta}-\bar{\mu}\right)}+1}dy\right]^2\,,\label{eq:PCFideal}
\end{align}
in the paramagnetic case. The Fermi--hole potential is simply evaluated by essentially inverting the Ornstein--Zernike equation within the HNC approximation\cite{DharmawardanaPRL2000,PerrotPRB2000,DharmawardanaIJQC2012},
\begin{align}
h_{\mathrm{HF}}^{\mathrm{uu}}(x)&=c_{\mathrm{HF}}^{\mathrm{uu}}(x)+\frac{1}{3\pi^2}\int c_{\mathrm{HF}}^{\mathrm{uu}}(x')h_{\mathrm{HF}}^{\mathrm{uu}}(|\boldsymbol{x}-\boldsymbol{x}'|)d^3x'\,,\label{OZequation}\\
g_{\mathrm{HF}}^{\mathrm{uu}}(x)&=\exp\left[-\beta\mathcal{P}(x)+h_{\mathrm{HF}}^{\mathrm{uu}}(x)-c_{\mathrm{HF}}^{\mathrm{uu}}(x)\right]\,,\label{HNCclosure}
\end{align}
where the non-interacting direct correlation function $c_{\mathrm{HF}}^{\mathrm{uu}}(x)$ is obtained from the Ornstein--Zernike equation and then the Fermi--hole potential $\beta\mathcal{P}(x)$ is obtained from the HNC closure. The idea of an effective potential description of the non-interacting electron gas originates from an early work by Lado\cite{LadoJCP1967}. The emerging total potential approximately considers both exchange and diffraction effects and is then used to solve the Ornstein--Zernike equation within the HNC closure for a spin-up/-down binary mixture\cite{DharmawardanaPRL2000,PerrotPRB2000}. The classical mapping method owes its popularity to the fact that the emerging interacting PCFs strictly respect non-negativity\cite{DornheimPhysRep2018}, in stark contrast to the dielectric formalism where a negative on--top PCF, $g(0)$, is rather unavoidable for most advanced schemes at reasonable coupling\cite{ToliasJCP2021,ToliasPRB2024}. For completeness, we note the Dufty \& Dutta efforts to introduce a more rigorous quantum--to--classical mapping based on the grand canonical ensemble\cite{DuftyPRE2013,DuttaPRE2013} as well as the Lue \& Wu efforts to go beyond the HNC approximation by introducing a bridge function\cite{LiuJCP2013}. It is also worth pointing out that a very accurate parametrization of the classical Coulomb bridge function has been recently reported, based on near-exact bridge function extractions from specially designed Molecular Dynamics simulations\cite{LuccoCastelloPRE2022}.

\subsection{Non-interacting static structure factor and the self-consistent dielectric formalism}\label{sec:dielectricformalism}

The self-consistent dielectric formalism was crystallized nearly sixty years ago in the classic paper by Singwi, Tosi, Land \& Sj\"olander\cite{SingwiPR1968}. Indicative of the longevity and versatily of the formalism is the fact that novel dielectric schemes of considerable complexity are still being proposed\cite{TanakaJCP2016,DornheimPRB2020,LuccoCastelloEPL2022,ToliasJCP2023}. However, the methodology for the solution of any dielectric scheme for the finite temperature interacting UEG (or any finite temperature quantum system) is based on the seminal works of Tanaka \& Ichimaru, who advocated the evaluation of the SSF through the Matsubara series expansion of Eq.(\ref{eq:MatsubaraSeries})\cite{TanakaJPSJ1986,TanakaPRA1985}. In that case, regardless of the dielectric scheme, the Matsubara summation of Eq.(\ref{eq:MatsubaraSeries}) is slowly converging, especially at high degeneracy. A speed up of the convergence is essential for the reduction of the computational cost of advanced dielectric schemes that feature dynamic local field corrections and address the strong coupling regime. This can be achieved by adding \& subtracting the Lindhard density response inside the infinite series and by benefitting from the knowledge of the non-interacting SSF. This mathematical trick was also proposed by Tanaka \& Ichimaru\cite{TanakaJPSJ1986,TanakaPRA1985}. To be more specific, let us briefly introduce the building blocks of the theory. The dielectric formalism combines the constitutive relation for the linear density response function, which defines the dynamic local field correction (LFC) $G(\boldsymbol{k},\omega)$,
\begin{align}
\chi(\boldsymbol{k},\omega)=\frac{\chi_0(\boldsymbol{k},\omega)}{1-U(\boldsymbol{k})\left[1-G(\boldsymbol{k},\omega)\right]\chi_0(\boldsymbol{k},\omega)}\,,\label{eq:densityresponseDLFC}
\end{align}
where $U(\boldsymbol{k})=4\pi{e}^2/k^2$ for Coulomb interactions, with a complicated expression for the LFC as a functional of the SSF, whose form depends on the dielectric scheme,
\begin{align}
G(\boldsymbol{k},\omega)\equiv{G}[S](\boldsymbol{k},\omega)\,.\label{eq:functionalclosure}
\end{align}
The above ingredients, Eqs.(\ref{eq:densityresponseDLFC},\ref{eq:functionalclosure}), together with the Matsubara summation for the SSF, Eq.(\ref{eq:MatsubaraSeries}), lead to a nonlinear functional equation of the type
\begin{align}
S(\boldsymbol{k})=-\frac{1}{{n}\beta}\displaystyle\sum_{l=-\infty}^{\infty}\frac{\widetilde{\chi}_0(\boldsymbol{k},\imath\omega_l)}{1-U(\boldsymbol{k})\left[1-{G}[S](\boldsymbol{k},\imath\omega_l)\right]\widetilde{\chi}_0(\boldsymbol{k},\imath\omega_l)}\,,\label{eq:functionalequation}
\end{align}
that should be numerically solved for the SSF. Adding \& substracting $\widetilde{\chi}_0(\boldsymbol{k},\imath\omega_l)$ inside the Matsubara summation, after some algebra, one obtains
\begin{align}
S(\boldsymbol{k})=S_{\mathrm{HF}}(\boldsymbol{k})-\frac{1}{{n}\beta}\displaystyle\sum_{l=-\infty}^{\infty}\frac{\widetilde{\chi}_0(\boldsymbol{k},\imath\omega_l)}{\left\{U(\boldsymbol{k})\left[1-{G}[S](\boldsymbol{k},\imath\omega_l)\right]\widetilde{\chi}_0(\boldsymbol{k},\imath\omega_l)\right\}^{-1}-1}\,.\label{eq:functionalequationconv}
\end{align}
In practice, the transformed Matsubara series, Eq.(\ref{eq:functionalequationconv}), converges much faster than the original Matsubara series, Eq.(\ref{eq:functionalequation}).

\subsection{Non-interacting static structure factor and thermodynamics}\label{sec:EOS}

There is a long history of theoretical investigations focused on the equation of state of the finite temperature UEG\cite{DornheimPhysRep2018,GrothCPP2017}. Modern parameterizations of the exchange-correlation free energy $\widetilde{f}_{\mathrm{xc}}(r_{\mathrm{s}},\Theta)$ or the interaction energy $\widetilde{u}_{\mathrm{int}}(r_{\mathrm{s}},\Theta)$ of the finite temperature UEG are based on PIMC simulations\cite{KarasievPRL2014,GrothPRL2017,KarasievPRL2018}. Here the $\,\widetilde{}\,$ symbol signifies that the energies are normalized by the Hartree energy $e^2/a_{\mathrm{B}}$ with $a_{\mathrm{B}}$ the first Bohr radius. The closed form expressions are constructed in a manner that respects the known near-exact ground state limit $\Theta\to0$ as determined by quantum Monte Carlo simulations, the known near-exact classical limit $\Theta\to\infty$ as determined by classical Molecular Dynamics or classical Monte Carlo simulations and the exact non-interacting limit $\lim_{r_{\mathrm{s}}\to0}\widetilde{u}_{\mathrm{int}}(r_{\mathrm{s}},\Theta)=-\alpha_{\mathrm{HF}}(\Theta)/r_{\mathrm{s}}$, where $r_{\mathrm{s}}=d/a_{\mathrm{B}}$ is the quantum coupling parameter and $\alpha_{\mathrm{HF}}$ is the so-called Hartree-Fock coefficient\cite{DornheimPhysRep2018}. This non-zero non-interacting limit is a direct consequence of exchange effects. The interaction energy of the UEG is given by the equivalent expressions\cite{DornheimPhysRep2018}
\begin{align}
\widetilde{u}_{\mathrm{int}}(r_{\mathrm{s}},\Theta)&=\frac{2}{3}\frac{1}{\pi\lambda{r}_{\mathrm{s}}}\int_0^{\infty}\left[g(x;r_{\mathrm{s}},\Theta)-1\right]xdx\,,\label{eq:interactiongen1}\\
\widetilde{u}_{\mathrm{int}}(r_{\mathrm{s}},\Theta)&=\frac{1}{\pi\lambda{r}_{\mathrm{s}}}\int_0^{\infty}\left[S(x;r_{\mathrm{s}},\Theta)-1\right]dx\,,\label{eq:interactiongen2}
\end{align}
where the numerical coefficient $\lambda=1/(dq_{\mathrm{F}})$ has been introduced. Let us consider the non-interacting limit in the paramagnetic case by substituting in Eq.(\ref{eq:interactiongen2}) for the non-interacting SSF via Eq.(\ref{idealSSFform3}). This leads to the expression
\begin{align}
\widetilde{u}_{\mathrm{int}}(r_{\mathrm{s}},\Theta)&=-\frac{3\Theta}{4\pi\lambda{r}_{\mathrm{s}}}\int_0^{\infty}\frac{1}{x}\int_0^{\infty}dy\frac{y}{\exp{\left(\frac{y^2}{\Theta}-\bar{\mu}\right)}+1}\ln{\left\{\left|\frac{1+\exp{\left[\bar{\mu}-\frac{(y-x)^2}{\Theta}\right]}}{1+\exp{\left[\bar{\mu}-\frac{(y+x)^2}{\Theta}\right]}}\right|\right\}}dx\,.\label{eq:interactiongen3}
\end{align}
Therefore, the Hartree-Fock coefficient is exactly defined by
\begin{align}
\alpha_{\mathrm{HF}}(\Theta)&=\frac{3\Theta}{4\pi\lambda}\int_0^{\infty}\frac{1}{x}\int_0^{\infty}dy\frac{y}{\exp{\left(\frac{y^2}{\Theta}-\bar{\mu}\right)}+1}\ln{\left\{\left|\frac{1+\exp{\left[\bar{\mu}-\frac{(y-x)^2}{\Theta}\right]}}{1+\exp{\left[\bar{\mu}-\frac{(y+x)^2}{\Theta}\right]}}\right|\right\}}dx\,.\label{eq:interactiongen4}
\end{align}
It is noted that the ground state limit of the Hartree-Fock coefficient can be evaluated exactly. Substitution of Eq.(\ref{eq:PCFidealground}) into Eq.(\ref{eq:interactiongen1}) or substitution of Eq.(\ref{idealSSFground}) into Eq.(\ref{eq:interactiongen2}) yield the exact result
\begin{align}
\alpha_{\mathrm{HF}}(\Theta=0)=\frac{3}{4\pi\lambda}\,.\label{eq:interactiongen5}
\end{align}
It is often not discussed that the high temperature limit of the Hartree-Fock coefficient can also be evaluated exactly. Substitution of Eq.(\ref{eq:PCFidealhigh}) into Eq.(\ref{eq:interactiongen1}) or substitution of Eq.(\ref{idealSSFhighT}) into Eq.(\ref{eq:interactiongen2}) yield the exact result
\begin{align}
\alpha_{\mathrm{HF}}(\Theta\to\infty)&=\frac{1}{3\pi\lambda\Theta}\,.\label{eq:interactiongen6}
\end{align}
In practice, the numerical evaluation of the Hartree-Fock coefficient is fitted to a Pade-like approximation that respects the exact ground state and exact high temperature limits. Such an expression was originally proposed by Perrot \& Dharma-wardana and reads as\cite{PerrotPRA1984}
\begin{align}
\alpha_{\mathrm{HF}}(\Theta)\simeq\frac{1}{\pi\lambda}\tanh{\left(\frac{1}{\Theta}\right)}\frac{0.75+3.04363\Theta^2-0.09227\Theta^3+1.7035\Theta^4}{1+8.31051\Theta^2+5.1105\Theta^4}\,,\label{eq:interactiongen7}
\end{align}
where we emphasize that $0.75=3/4$ and that $1.7035/5.1105=1/3$. It is worth pointing out that the Perrot \& Dharma-wardana parameterization of the Hartree-Fock coefficient is highly accurate, in spite of being based on a truncated series evaluation and only for $0.1\leq\Theta\leq12$. An improvement of Eq.(\ref{eq:interactiongen7}) is possible by numerically evaluating Eq.(\ref{eq:interactiongen4}) in a wider range of degeneracies and by expanding the Pade approximant to higher orders of the degeneracy temperature, but it is probably rather inconsequential at a practical level.

\subsection{Non-interacting imaginary time correlation function and the self-consistent dielectric formalism}\label{sec:dielectricformalismimag}

As aforementioned, PIMC simulations provide direct access to the ITCF of finite temperature quantum systems\cite{CeperleyRMP1995}. In order to acquire the quasi-exact DSF, one needs to invert the two-sided Laplace transform of Eq.(\ref{eq:ITCFdef}). Unfortunately, this is a well-known ill-posed problem with respect to the unavoidable Monte Carlo error bars\cite{JarrellPhysRep1996,ShaoPhysRep2023}. Despite the development of several inversion methods, the analytic continuation challenge remains. It has been recently argued that a paradigm shift from the real frequency domain of DSFs to the imaginary-time domain of ITCFs would practically circumvent the analytic continuation problem\cite{DornheimPRB2023,DornheimPoPRev2023,DornheimMRE2023,DornheimPTRSA2023}. The domains are complementary and encode the same physics given the uniqueness of two-sided Laplace transforms\cite{DornheimPoPRev2023,DornheimMRE2023}. In the experimental front, this shift has already led to the development of a high accuracy model-free temperature diagnostic based on X-ray Thomson scattering (XRTS) measurements\cite{DornheimNat2022,DornheimPoP2023}, a formally exact f-sum rule based approach for the normalization of XRTS spectra\cite{DornheimSciRep2024}, the direct extraction of the Rayleigh weight from XRTS measurements without the need for any modelling assumptions or computer simulations\cite{dornheim2024modelfreerayleighweightxray} and the first-principle PIMC-based analysis of XRTS measurements for strongly compressed beryllium obtained at the National Ignition Facility\cite{dornheim2024unravelingelectroniccorrelationswarm}. In the theoretical front, this shift inspired the derivation of the Fourier--Matsubara series expansion that connects the ITCF with the linear density response function\cite{ToliasJCP2024}, Eq.(\ref{eq:FourierMatsubaraSeries}), as well as recent efforts to extract and parameterize the dynamic local field correction evaluated at the Matsubara frequencies\cite{DornheimPRB2024,DornheimEPL2024,dornheim2024shortwavelengthlimitdynamic}. At this point, it should be emphasized that the self-consistent dielectric formalism gives direct access not only to the SSF but also to the density response function evaluated at the imaginary Matsubara frequencies. This implies that the dielectric formalism is tailor-made for the imaginary-time domain provided that the appropriate theoretical tools are utilized. As very recently discussed\cite{ToliasJCP2024}, for any dielectric scheme, in order to obtain the ITCF in the imaginary--time domain from real frequency domain calculations, one needs to use the SSF for the calculation of the complex dynamic LFC in the frequency domain via the complicated closure functional of Eq.(\ref{eq:functionalclosure}), to employ the LFC to evaluate the complex dynamic density response function in the frequency domain via the constitutive relation of Eq.(\ref{eq:densityresponseDLFC}), to use the imaginary part of the density response function to compute the DSF through the FDT of Eq.(\ref{eq:quantumFDT}) and to apply the two-sided Laplace transform of Eq.(\ref{eq:ITCFdef}) that involves a complicated frequency integration over the longitudinal collective modes. On the other hand, for any dielectric scheme, in order to obtain the ITCF in the imaginary time domain from Matsubara frequency domain calculations, one simply needs to substitute for the Matsubara density response functions in the Fourier--Matsubara series expansion of Eq.(\ref{eq:FourierMatsubaraSeries}). The only possible complication concerns the slow convergence of the Matsubara summation of Eq.(\ref{eq:FourierMatsubaraSeries}), especially at high degeneracy. This implies that the Fourier--Matsubara series should be truncated at very high values of the Matsubara order leading to an increase in the computational cost. As already discussed in section \ref{sec:dielectricformalism}, a significant acceleration can be achieved by adding \& subtracting $\widetilde{\chi}_0(\boldsymbol{k},\imath\omega_l)e^{-\imath\hbar\omega_l\tau}$ within the infinite series and by benefitting from the knowledge of the non-interacting ITCF. This converts
\begin{align}
F(\boldsymbol{k},\tau)=-\frac{1}{{n}\beta}\displaystyle\sum_{l=-\infty}^{\infty}\frac{\widetilde{\chi}_0(\boldsymbol{k},\imath\omega_l)}{1-U(\boldsymbol{k})\left[1-{G}[S](\boldsymbol{k},\imath\omega_l)\right]\widetilde{\chi}_0(\boldsymbol{k},\imath\omega_l)}e^{-\imath\hbar\omega_l\tau}\,,\label{eq:functionalequationimag1}
\end{align}
to
\begin{align}
F(\boldsymbol{k},\tau)=F_{\mathrm{HF}}(\boldsymbol{k},\tau)-\frac{1}{{n}\beta}\displaystyle\sum_{l=-\infty}^{\infty}\frac{\widetilde{\chi}_0(\boldsymbol{k},\imath\omega_l)}{\left\{U(\boldsymbol{k})\left[1-{G}[S](\boldsymbol{k},\imath\omega_l)\right]\widetilde{\chi}_0(\boldsymbol{k},\imath\omega_l)\right\}^{-1}-1}e^{-\imath\hbar\omega_l\tau}\,.\label{eq:functionalequationimag2}
\end{align}

\section{Future work}\label{sec:future}

In recent years, there has been increasing interest in the nonlinear electronic density response of warm dense matter\cite{DornheimPoPRev2023,MoldabekovJCTC2022}. It has been observed that nonlinear effects strongly depend on the level of degeneracy and that nonlinear effects are non-negligible for various experimentally accessible conditions\cite{DornheimPRL2020}. Nevertheless, it remains unclear whether existing diagnostics with particle beams and X-rays can be effectively tuned to reliably investigate nonlinear observables. Extensive nonlinear investigations are now available either through PIMC simulations of the harmonically perturbed UEG (extraction of the static nonlinear density response at a specific wavenumber based on its definition)\cite{DornheimPRR2021,DornheimCPP2021,DornheimJPSJ2021,DornheimJCP2023b} or through PIMC simulations of the equilibrium UEG (extraction of the static nonlinear density response at all wavenumbers based on the integration of higher-order ITCFs)\cite{DornheimJCP2021,DornheimCPP2022}. Moreover, arbitrarily higher-order generalizations of the ideal density response are readily available\cite{ToliasEPL2023,MikhailovPRL2014}. In addition, higher-order connections between PIMC-extractable ITCFs and DSFs have been already established\cite{vorberger2024greensfunctionperspectivenonlinear,DornheimJCP2021}. Furthermore, the general relations of response theory (Kubo formula, fluctuation dissipation theorem, Matsubara summation) are also generalizable to higher-orders\cite{HuPRB1988,EcheniquePRA1986}. Thus, the present article sets the stage for future investigations of the higher-order density correlations of the non-interacting finite temperature electron gas that are invariably connected with the nonlinear density response\cite{vorberger2024greensfunctionperspectivenonlinear}. It is also worth emphasizing that $s$-reduced correlation functions are available for the non-interacting electron gas for any dimensionality and particle number, but only in the ground state\cite{TanakaJPSJ2011}.

\section*{Acknowledgements}

This work was partially supported by the Center for Advanced Systems Understanding (CASUS), financed by Germany’s Federal Ministry of Education and Research (BMBF) and the Saxon state government out of the State budget approved by the Saxon State Parliament. This work has received funding from the European Research Council (ERC) under the European Union’s Horizon 2022 research and innovation programme (Grant agreement No. 101076233, "PREXTREME"). Views and opinions expressed are however those of the authors only and do not necessarily reflect those of the European Union or the European Research Council Executive Agency. Neither the European Union nor the granting authority can be held responsible for them.

\bibliography{main_bib}

\end{document}